\documentclass[sigconf,nonacm]{acmart}

\usepackage{booktabs}
\usepackage{listings}
\usepackage{xcolor}

\setcopyright{none}
\renewcommand\footnotetextcopyrightpermission[1]{}

\begin{document}

\title{Natural Language Access to Domain-Specific Metadata:\\A Reusable Framework for LLM Query Generation}

\author{Blake G. Fitch}
\email{blake.fitch@tuebingen.mpg.de}
\affiliation{%
  \institution{Max Planck Institute for Biological Cybernetics}
  \city{T\"ubingen}
  \country{Germany}}

\author{Cato Elia Kurtz}
\affiliation{%
  \institution{Max Planck Institute for Biological Cybernetics}
  \city{T\"ubingen}
  \country{Germany}}

\begin{abstract}
Researchers need to answer ad-hoc questions about the contents of domain-specific archives but often lack the expertise to write structured queries on the metadata.
We show that when domain vocabulary and semantics are captured in a well-designed Web Ontology Language (OWL) ontology, Large Language Models (LLMs) can generate accurate structured queries zero-shot, that is, without task-specific fine-tuning examples, retrieval augmentation, or multi-agent orchestration.
We present the Natural Language Knowledge Graph Query (NLKGQ) system, a framework and development process that enables natural language access to metadata in such archives.
The framework includes a web interface that helps researchers pose natural language questions, which a domain-agnostic harness translates to SPARQL via an LLM and executes against a knowledge graph.
The development process begins with capturing domain vocabulary and semantics in a formal OWL ontology.
Domain-specific code then extracts metadata from archive sources and imports it into a knowledge graph defined by the ontology.
Both the framework and process are designed to support reuse across multiple domains.
In this work, we demonstrate the system for metadata derived from a large-scale neuroimaging research archive, evaluating performance across multiple LLMs and ontology representations.
The best configurations achieve 100\% accuracy on a 21-question competency and regression test set developed with domain experts.
An ablation study across eight ontology representations reveals that readable entity names and semantic annotations are the dominant factors in accuracy, more significant than model choice or prompt engineering.
We also compare SPARQL to an auto-generated SQL database as query backends, showing that OWL's structural features provide a substantial advantage over SQL DDL for LLM-driven query generation.
A notable consideration for our demonstration domain is a requirement to run local LLMs on modest institutional hardware in support of privacy concerns for human subject data.
\end{abstract}


\maketitle

\section{Introduction}
\label{sec:intro}

Research facilities, enterprises, and government agencies accumulate large archives of domain-specific data with rich metadata, for example subject demographics, acquisition parameters, experimental configurations, and provenance records.
Querying this metadata ad-hoc can require writing structured queries in SQL or SPARQL, which in turn requires both knowledge of the query language and the domain vocabulary.
Many researchers lack the expertise to effectively access the information they need without assistance.

Large language models (LLMs) can generate structured queries from natural language (NL), but their accuracy depends on how the underlying data schema is presented.
On simple schemas like Spider~\cite{yu2018spider}, top models exceed 90\% accuracy, but on real enterprise schemas with hundreds of tables, GPT-4o drops to 0\%~\cite{chen2024beaver}.
Sequeda et al.~\cite{sequeda2023benchmark} find that reformulating the same enterprise questions over a knowledge graph raises accuracy from 16\% to 54\%.
How domain knowledge is structured matters as much as which LLM model is used.

In this paper, we present an ontology-first process for enabling NL access to domain-specific metadata, a generic framework of tools to facilitate deployment, and measured results searching metadata on a large MRI image archive.
The process begins with capturing the domain vocabulary and semantics in a formal OWL ontology~\cite{owl2} using readable entity names and semantic annotations.
A domain-specific Extract-Transform-Load (ETL) pipeline maps archive metadata to the ontology, producing a knowledge graph (KG).
Using the KG stored in RDF Turtle format, we load both a SPARQL database and, via automatic transformation, an SQL database, enabling a systematic comparison of both as targets for LLM-generated queries with equivalent data and NL questions.
An LLM generates correct queries against either backend, zero-shot, with the full ontology or derived schema provided in full, in the system prompt.

The domain-specific ontology, competency and regression test cases, and a domain-specific prompt rider evolve together through iterative testing. During this process, the vocabulary and semantics of the domain are refined in increasing detail, progressively eliminating confusion and inaccuracies that plague even expert human communication.

We demonstrate this process on an MRI neuroimaging research archive~\cite{fitch2022mrdata} containing metadata for 80 active studies with 2,000 MRI experiments.
Since this archive contains human subject data, privacy regulations (GDPR) require that all processing remain on institutional infrastructure rather than external LLM services.
A practical objective is therefore to find the smallest model that achieves acceptable accuracy, keeping hardware costs manageable for institutions that may have modest GPU resources.
We evaluate across 8 locally deployed LLM variants (8B to 35B parameters, including quantized and mixture-of-experts variants), showing that the best model achieves 100\% accuracy on SPARQL and 57\% on auto-generated SQL, with no fine-tuning, no retrieval augmentation, and no multi-agent orchestration~\cite{zhang2024gail,zhao2025agentigraph,zhao2025cyberbot,walter2025grasp}.
Stripping ontology annotations degrades SPARQL accuracy by 19 percentage points and SQL by 14, revealing the importance of semantic annotations for LLM-driven query generation.

\textbf{Contributions:}
\begin{enumerate}
    \item The NLKGQ framework and ontology-first development process for enabling NL access to domain-specific metadata, including ontology design principles, a KG builder pattern, and a reusable query server with web interface, demonstrated on neuroimaging research data
    \item A systematic evaluation of LLM-driven SPARQL generation across 8 models, 8 ontology representations, and 768 configurations (over 16,000 runs), showing that ontology design is a dominant factor in accuracy
    \item A generic OWL KG-to-SQL schema generator and data transformer enabling a systematic comparison of text-to-SPARQL vs text-to-SQL for the same NL questions
    \item Analysis of the structural advantages OWL provides over SQL Data Definition Language (DDL) for LLM-based query generation
\end{enumerate}

\section{Related Work}
\label{sec:related}

\subsection{Knowledge Graph Question Answering}

Traditional Knowledge Graph Question Answering (KGQA) systems combine entity linking with relation prediction, mapping NL to structured queries through classification~\cite{usbeck2018qald9,hogan2021kg}.
Recent work applies LLMs to this problem through several strategies.

Fine-tuning approaches train models on SPARQL examples.
SGPT~\cite{rony2022sgpt} applied a pre-trained generative model, combined with knowledge graph embeddings, to SPARQL generation.
Zhang et al.~\cite{zhang2024gail} use GAIL (Generative Adversarial Imitation Learning) to fine-tune LLMs for low-resource KGQA, where the LLM acts as a generator producing SPARQL that a discriminator evaluates against expert demonstrations.

Multi-agent and pipeline approaches add orchestration layers.
AGENTiGraph~\cite{zhao2025agentigraph} deploys intent classification, task planning, and automatic knowledge integration across multiple agents, achieving 95\% classification accuracy on a 3,500-query benchmark.
CyberBOT~\cite{zhao2025cyberbot} combines RAG with an ontology-based verification layer that constrains LLM outputs post-hoc.

In-context learning approaches provide examples at inference time.
D'Abramo et al.~\cite{dabramo2025icl_sparql} show that ICL with retrieved SPARQL examples can match fine-tuned models on KGQA benchmarks, though their approach requires example retrieval infrastructure.
GRASP~\cite{walter2025grasp} uses the LLM to explore the knowledge graph at query time, searching for relevant IRIs and literals, achieving state-of-the-art results on Wikidata in a zero-shot setting.

Rhizomer-LLM~\cite{paslavska2026rhizomerllm} lowers the barrier to SPARQL for non-expert users with an LLM-assisted exploratory interface over cloud APIs (Gemini, DeepSeek, Llama), evaluated on the BESDUI task-completion benchmark.
Because BESDUI measures whether a task was completed rather than whether the returned answer is correct, since, in the authors' words, semantic correctness cannot be automatically verified, it is complementary to our evaluation, which reports execution-level result correctness against a reference query for every competency question.

All of these approaches build complexity to cope with ontologies they cannot change, typically public KGs like Wikidata~\cite{vrandecic2014wikidata} or DBpedia, where opaque identifiers provide few semantic hints to the LLM.
Our approach differs in assuming the ontology is under our control.
Domain-specific ontologies are typically smaller in vocabulary than public KGs, but may describe large volumes of data with a compact set of classes and properties.

\subsection{Text-to-SQL and Schema Representation}

Text-to-SQL has seen rapid progress on benchmarks like Spider~\cite{yu2018spider}, but performance depends on schema simplicity.
BEAVER~\cite{chen2024beaver} shows that GPT-4o achieves 0\% execution accuracy on real enterprise schemas with hundreds of tables, with failures traced to schema retrieval and column mapping.
Sequeda et al.~\cite{sequeda2023benchmark} show that knowledge graphs provide a structural advantage, raising accuracy from 16\% (SQL) to 54\% (SPARQL) on the same enterprise questions.
Rajkumar et al.~\cite{rajkumar2022schema} show that schema representation, how table and column names are presented, directly affects LLM accuracy for SQL generation, paralleling our findings for SPARQL.
Wretblad et al.~\cite{wretblad2024column_descriptions} show that adding column descriptions to SQL schemas improves text-to-SQL accuracy by over 20\% on columns with uninformative names, validating the importance of annotations.

\subsection{Text-to-SPARQL}

LLM-based approaches to SPARQL generation began with sequence-to-sequence models~\cite{soru2017sparql}.
Chain-of-thought prompting~\cite{zahera2024cotsparql} and complex benchmarks like Spider4SPARQL~\cite{kosten2023spider4sparql} have advanced the field, but most work targets public KGs with opaque identifiers.

Giuliani et al.~\cite{giuliani2026sparql} propose heuristics for designing LLM-friendly ontologies, such as preferring semantically expressive labels over cryptic identifiers, drawn from experience refining evaluation ontologies for two zero-shot SPARQL benchmarks.
Their appendix states our controlled-semantics thesis as design advice, without an ablation isolating naming from other confounded changes: their one supporting data point, GPT-3.5 improving from 0.08 to 0.60, combines renaming, added axioms, and query filtering in a single step.
Our ablation (Table~\ref{tab:ontology_results}) provides the controlled measurement behind such guidance: holding model, prompt, and benchmark fixed, ontology representation alone accounts for a 90-point spread in SPARQL accuracy, from 100\% (default) to 10\% (abstract-graph).

SparqLLM~\cite{arazzi2025sparqllm} retrieves from 360 hand-authored query templates to construct SPARQL via retrieval-augmented generation; removing the template retrieval drops their reported relaxed-match accuracy from 66.7\% to 55.3\%, indicating that curated templates, not the LLM's zero-shot capability, account for most of their accuracy.
Our approach requires no per-question template curation: the ontology alone, supplied once in the system prompt, provides the semantic scaffolding an LLM needs to generate correct queries for a fixed domain.

Vejvar and Fujimoto~\cite{vejvar2026spider4ssc} compare SPARQL, SQL, and Cypher generation on a shared benchmark and find SPARQL trails SQL under terse, token-dense schema linearizations (17.2\% vs 29.7\% execution accuracy with fine-tuned uT5 models, including for a pretrained model tested in their appendix); they identify richer schema representations as future work.
We test that condition directly: with the full annotated OWL ontology in context, SPARQL accuracy exceeds auto-generated SQL (100\% vs 57\%), suggesting that schema and ontology richness, not the query language itself, determines which backend an LLM can query more reliably.

Closest to our work, Rasheed and Aguado~\cite{rasheed2025domain_sparql} investigate prompt augmentation formats for domain-specific KGs, testing reduced and condensed ontology representations.
However, they take the existing KG ontology as given and focus on fitting it into the prompt, rather than designing the ontology for LLM consumption from the start.
Our work focuses on domain-specific ontologies of finite size that fit in the LLM context window.
Our measurements show that departing from expressive naming of ontological elements hurts accuracy.

\subsection{Our Focus}

Prior work treats the ontology or database schema as a given and focuses on improving query generation.
To our knowledge, no prior work addresses the end-to-end process: how to capture domain vocabulary in a formal ontology, build a knowledge graph from domain metadata via ETL, and enable LLM-driven NL querying, all as a reusable technique for domain practitioners.

Furthermore, while Sequeda et al.~\cite{sequeda2023benchmark} compare LLM accuracy on SQL vs SPARQL across different schemas and datasets, no prior work compares SPARQL and SQL as query backends when both are derived from the same authoritative ontology, with the same readable naming, the same data, and the same LLM models.
We provide this controlled comparison by automatically generating a relational schema from the domain ontology and evaluating both backends on the same competency questions.

\section{Production Framework with Demonstration Domain}
\label{sec:framework}

Our contribution has two main areas: a \textit{development process} and a \textit{reusable framework}.
The development process is what a domain practitioner follows to enable NL access for a new domain: (1)~capture domain vocabulary in an OWL ontology, (2)~build a knowledge graph via ETL from domain data sources, and (3)~iterate the ontology, competency questions, and prompt rider together until accuracy is acceptable.
The reusable framework is the domain-agnostic infrastructure that supports this process: the NLKGQ query server, web application, and combinatorial test driver.
The framework also supports production deployment.
Both are described below, illustrated with our neuroimaging demonstration domain.

\subsection{Capturing Domain Vocabulary}
\label{sec:ontology_design}

The first step is formalizing what domain experts already know: the entities, their properties, their relationships, and what they mean.
We capture this in an OWL ontology~\cite{owl2,noy2001ontology101} following six design principles that cost nothing in formal expressiveness but measurably improve LLM accuracy:

\begin{enumerate}
    \item \textbf{Full English words}: \texttt{hasAcquisitionTime} rather than \texttt{acqT} or \texttt{AT}
    \item \textbf{Consistent naming pattern}: Properties follow \texttt{has<Property>} (e.g., \texttt{hasSubjectId}, \texttt{hasRepetitionTime})
    \item \textbf{Descriptive relationships}: \texttt{isSubjectOfExperiment} clearly indicates direction
    \item \textbf{Explicit domain and range}: Property definitions specify which classes they connect
    \item \textbf{Natural language annotations}: \texttt{rdfs:comment}, \texttt{rdfs:label}, and \texttt{skos:altLabel} provide context, readable names, and synonyms beyond the property URI
    \item \textbf{No opaque codes}: Meaningful URIs throughout; no arbitrary or auto-generated identifiers
\end{enumerate}

These are not cosmetic choices.
They are engineering decisions made during ontology development.
When building a new domain knowledge graph, readability costs little: the ontology must be designed regardless, and readable names are no harder to define than opaque ones.
These principles also align with the FAIR (Findable, Accessible, Interoperable, Reusable) data guidelines~\cite{wilkinson2016fair}, particularly Reusable: the ontology provides rich, domain-relevant metadata described with formal semantics, making both the data and the NL access layer reusable across tools and communities.

Consider a concrete example from our demonstration domain ontology (MRI Research Ontology, MRO):

\begin{lstlisting}
mro:hasHandedness a owl:DatatypeProperty ;
    rdfs:label "Handedness"@en ;
    rdfs:comment "Subject handedness: Right, Left,
        Ambidextrous, or Unknown. (MRO)"@en ;
    rdfs:domain mro:Subject ;
    rdfs:range xsd:string .
\end{lstlisting}

The property name, label, comment, domain, and range together allow the LLM to use this property correctly without any prior exposure to our ontology.
The comment tells the LLM what values to expect, enabling it to generate correct \texttt{FILTER} clauses.
Contrast this with Wikidata's \texttt{wdt:P552} (handedness): the identifier P552 carries no semantic information, so the mapping must be learned through fine-tuning, few-shot examples, or graph exploration.

Frontier LLMs can assist domain experts in formalizing their vocabulary into OWL, lowering the barrier to ontology creation for practitioners without semantic web expertise.

\begin{figure}[t]
\centering
\fbox{\includegraphics[width=0.97\columnwidth]{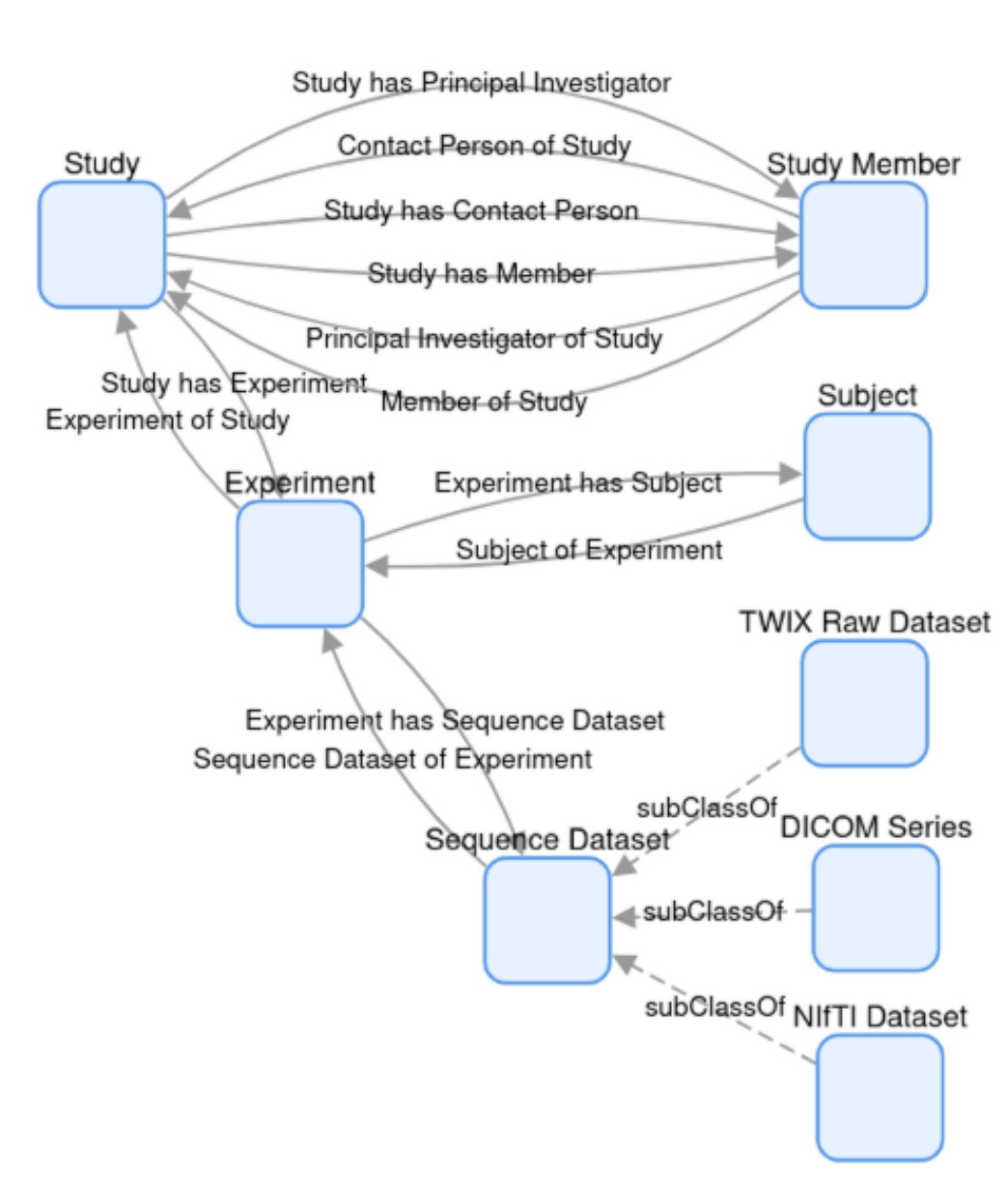}}
\caption{MRI Research Ontology (MRO) class diagram. Classes are connected by object properties (arrows).}
\label{fig:ontology}
\end{figure}

Figure~\ref{fig:ontology} shows the class and property structure of our demonstration domain ontology (MRO).

\subsection{ETL and Knowledge Graph Construction}
\label{sec:etl}

With the domain ontology defined, a domain-specific ETL pipeline collects metadata and populates a KG.
In our demonstration domain, metadata is extracted from DICOM headers, NIfTI sidecar files~\cite{gorgolewski2016bids,li2016dcm2niix}, Siemens TWIX raw data headers, an experiment registry, and a participant management system~\cite{mader2023castellum}.
Our Python pipeline uses rdflib~\cite{rdflib} to map extracted fields to MRO classes and properties, producing RDF triples.
The resulting KG contains approximately 10 million triples derived from a 180TB MRI archive.

Each new domain requires its own ETL logic, but the pattern is the same: identify metadata sources, write extraction code, map fields to ontology terms, generate RDF.
The ETL pipeline can also anonymize sensitive fields and compute aggregate properties (e.g., experiment duration from individual acquisition timestamps) during the transformation step.
Rather than relying on an OWL reasoner at query time, the ETL pipeline materializes all implied relationships as concrete RDF triples (e.g., generating both directions of inverse properties and explicit class memberships) during KG construction; this is sufficient for the OWL constructs we use.
The ontology guides what to extract and how to name it.

The resulting KG is loaded into a SPARQL server, in our case Apache Jena Fuseki~\cite{carroll2004jena}, where it can be queried using SPARQL 1.1~\cite{sparql11} via the web interface or HTTP API.
This keeps the SPARQL server lightweight and query latency predictable.

\subsection{Iterative Development}
\label{sec:iterative}

Ontology development is not a one-time design step.
The ontology, competency questions, and domain-specific prompt rider co-evolve through testing.
Running test cases against the system reveals gaps that feed back into all three:

\begin{itemize}
    \item A property name that confuses the LLM gets renamed in the ontology (e.g., adding \texttt{skos:altLabel} ``pseudonym'' to \texttt{hasSubjectId} when users say ``pseudonym'' instead of ``subject ID'')
    \item A recurring LLM mistake gets addressed by a new directive in the domain-specific prompt rider (e.g., ``return experiment IDs, not instance URIs'')
    \item A new user question that the system handles poorly becomes a new competency test case
\end{itemize}

This cycle is lightweight: each iteration is a small change to one of the three artifacts, followed by re-running the test suite.
It requires domain knowledge and comfort with iterative testing, but no machine learning, database, or SPARQL expertise.
The result is a progressively refined system where the ontology captures the domain vocabulary, the test cases capture the query patterns, and the rider captures the conventions that naming alone cannot express.

\subsection{NLKGQ Query Server and Web Application}
\label{sec:nlkgq}

The NLKGQ server receives an NL question and constructs a system prompt containing:
(1)~generic SPARQL generation instructions (output format, error handling, query patterns to prefer or avoid),
(2)~the complete domain ontology in OWL Turtle format, and
(3)~a small domain-specific rider with conventions particular to the knowledge graph (e.g., preferred identifier patterns, directives addressing recurring LLM mistakes).
The user's NL question is appended as the user message, and the LLM generates SPARQL.

Logic in the harness handles practical issues of the LLM response: stripping chain-of-thought \texttt{<think>} blocks, extracting SPARQL from markdown, fixing PREFIX declarations using the ontology as ground truth, and optionally retrying failed queries by sending the error back to the LLM.
Retrying failed queries is technically not zero-shot; however, it primarily helps smaller models recover from syntax errors, while the best-performing models rarely trigger it.

\begin{figure}[t]
\centering
\includegraphics[width=\columnwidth]{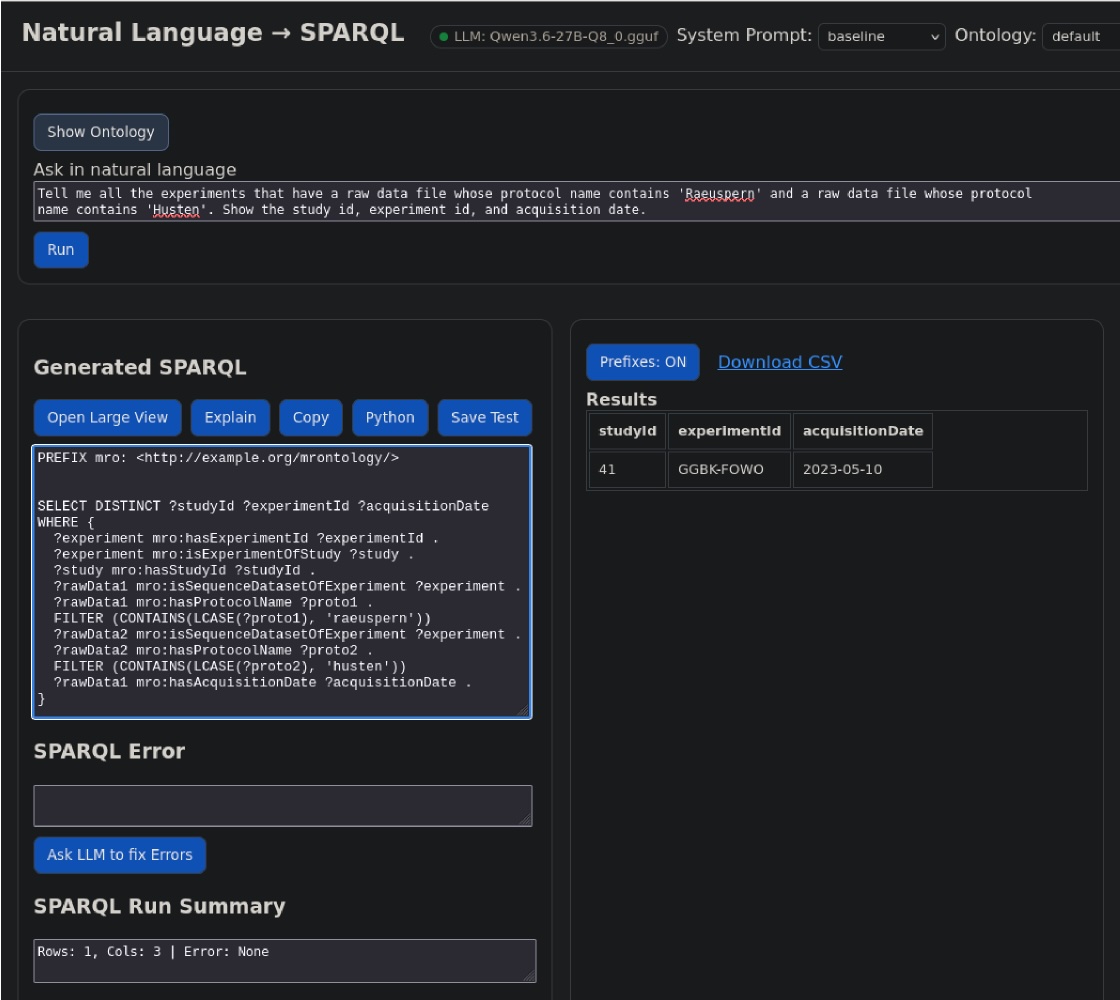}
\caption{The NLKGQ web interface. The user types a natural language question (top), the system generates and displays the SPARQL query (left), and shows the result table with prefix-compressed URIs (right).}
\label{fig:webapp}
\end{figure}

The web application provides an interface where end users type NL questions and receive query results, with the generated SPARQL shown for transparency.
Features include prefix compression that strips verbose RDF URIs from query results, displaying data with readable prefixed names instead (with a toggle for full URIs when needed), CSV download of results, an LLM-powered ``Explain'' button that describes what the generated SPARQL does in plain language, Python script export for rerunning queries outside the web interface, an interactive ontology graph visualization, and a ``Save Test Case'' function that packages a query with its results for use in the evaluation framework.
Displaying the generated query builds user trust and helps researchers learn which phrasings produce useful results.
The system is deployed at a neuroimaging research facility where neuroscience researchers with varying levels of query expertise can search their metadata directly.
Figure~\ref{fig:webapp} shows the web interface processing a natural language query.

\begin{figure*}[t]
\centering
\includegraphics[width=\textwidth]{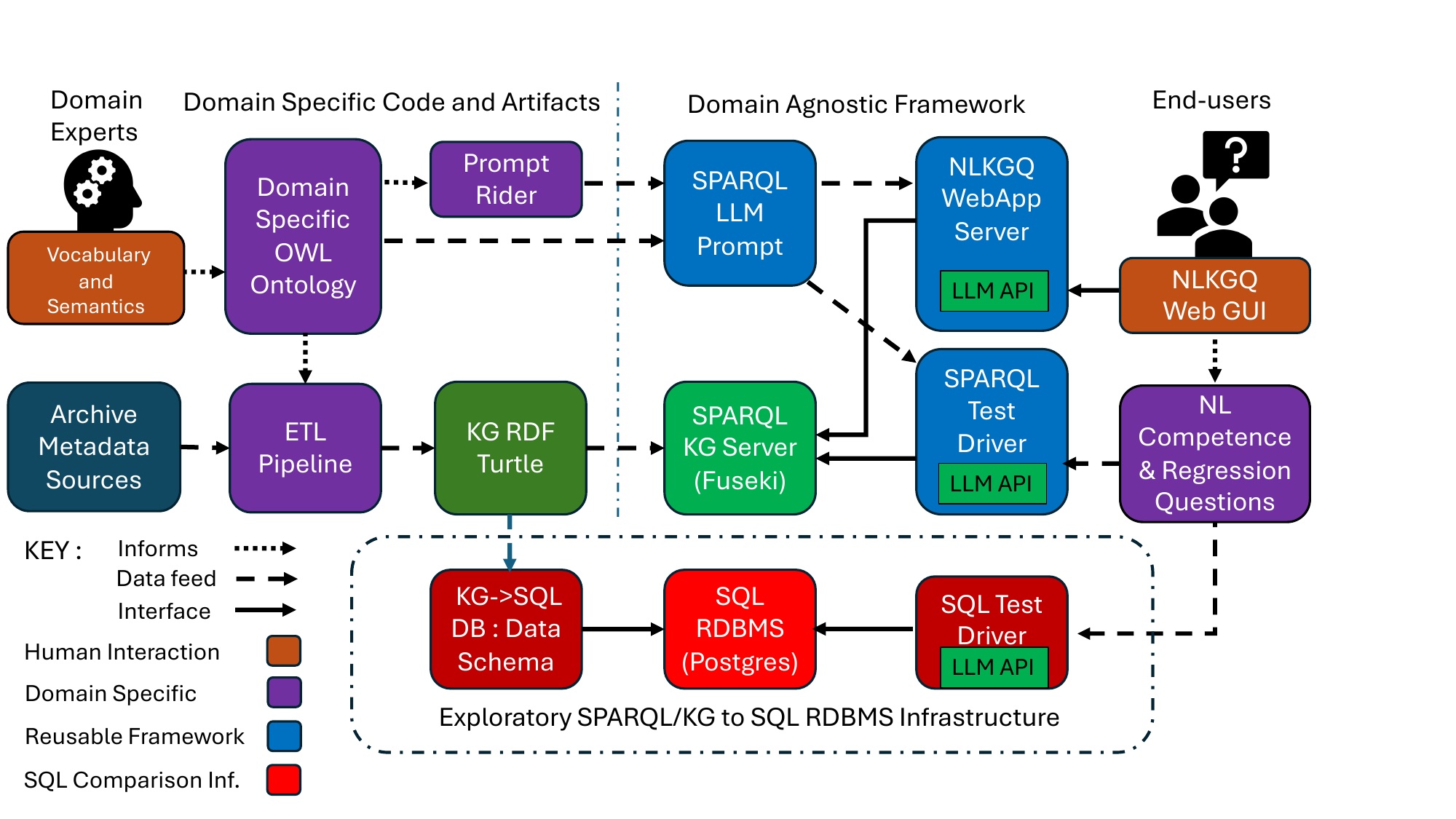}
\caption{Framework architecture showing production components (solid) and research infrastructure (dashed).}
\label{fig:architecture}
\end{figure*}

\section{Research Methods and Experimental Setup}
\label{sec:methods}

To rigorously evaluate our design decisions, we developed additional research infrastructure beyond the production system.
Some of this infrastructure also supports the iterative development process (Section~\ref{sec:iterative}), serving both production and research needs.

\subsection{Test Driver Framework}
\label{sec:testdriver}

The test driver is the foundation of both domain development and research evaluation.
Given a competency question (NL), a reference SPARQL query, and expected results, the driver calls the LLM, extracts the generated query, executes it against the appropriate backend, and compares results.
The driver supports combinatorial sweeps across the following parameters for each competency question:

\begin{itemize}
    \item \textbf{LLM model}: different model families, sizes, and quantization variants
    \item \textbf{Temperature}: controls generation randomness (e.g., 0.0, 0.2)
    \item \textbf{System prompt}: different prompt formulations and domain-specific riders
    \item \textbf{Ontology representation}: full Turtle, no-comments, compact, abstract, etc.
    \item \textbf{Query backend}: SPARQL or auto-generated SQL (with or without column comments)
    \item \textbf{Repetitions}: multiple runs per configuration to assess variability
    \item \textbf{Fix-retry policy}: number of LLM correction attempts on syntax errors
\end{itemize}

Results are logged to CSV with per-query timing, match status, and fix-retry outcomes.

A walker script orchestrates the driver across all test cases and configurations, with parallel execution across models.
For the SQL evaluation, a separate test driver executes against PostgreSQL with tolerant comparison (column name normalization, order-independent row matching, and URI prefix stripping) to handle differences between SQL output and SPARQL reference data.

The test driver is used both for iterative domain development (Section~\ref{sec:iterative}) and for the research evaluations described below.

\subsection{Virtual Ontology Representations}
\label{sec:virtual_ontology}

To measure which ontology features different LLMs rely on, we evaluate eight representations of the same ontology, generated automatically from the full OWL Turtle:

\begin{itemize}
    \item \textbf{default}: Full OWL Turtle with domain, range, comments, labels
    \item \textbf{no-comments}: Turtle with \texttt{rdfs:comment} and \texttt{rdfs:label} stripped
    \item \textbf{compact}: One-line-per-property listing class memberships
    \item \textbf{compact-typed}: Compact with datatype annotations retained
    \item \textbf{compact-grouped}: Properties grouped by domain class
    \item \textbf{raw-graph}: RDF triples in N-Triples format
    \item \textbf{abstract-dict}: Dictionary mapping generic keys to values
    \item \textbf{abstract-graph}: Graph with anonymized node/edge labels
\end{itemize}

These form an ablation study: moving from \textit{default} to \textit{abstract-graph} progressively strips semantic information, revealing which features the LLM relies on.

\subsection{Automatic Generation of SQL Database}
\label{sec:autosql}

We developed a generic OWL-to-SQL converter that reads the domain ontology and automatically derives a PostgreSQL 17 schema: each OWL class becomes a table, each datatype property becomes a column (typed from \texttt{rdfs:range}), and object properties become foreign keys (direction inferred from \texttt{owl:inverseOf}).
Ontology annotations (\texttt{rdfs:comment}, \texttt{rdfs:label}, \texttt{skos:altLabel}) are converted to SQL column comments.
The same KG data is loaded into the relational schema (6 tables, 182 columns for the largest table).

A system prompt analogous to the SPARQL prompt provides the Data Definition Language (DDL) schema (with or without comments) and instructs the LLM to generate SQL instead of SPARQL.
A SQL test driver evaluates the same competency questions against PostgreSQL using tolerant comparison to match SQL results against SPARQL reference data.

\subsection{Models and Hardware}
\label{sec:models}

We evaluate eight model variants from the Qwen3 family, spanning 8B to 35B parameters, including dense and mixture-of-experts (MoE) architectures, and quantized variants for production deployment:

\begin{table}[h]
\small
\caption{Models evaluated. All served locally. HW: A=AMD MI300A APU (HPC), N=4$\times$NVIDIA RTX5000 (institutional).}
\label{tab:models}
\begin{tabular}{lrlrll}
\toprule
Model & Params & Type & Ctx & Server & HW \\
\midrule
Qwen3-8B & 8B & Dense & 41K & vLLM & A \\
Qwen3-14B & 14B & Dense & 41K & vLLM & A \\
Qwen3-Coder-30B-A3B & 30B/3B & MoE & 64K & vLLM & A \\
Qwen3.6-27B & 27B & Dense & 64K & vLLM & A \\
Qwen3.6-27B-FP8 & 27B & Dense/FP8 & 64K & vLLM & A \\
Qwen3.6-27B-Q8\_0 & 27B & Dense/Q8 & 32K & llama.cpp & N \\
Qwen3.6-35B-A3B & 35B/3B & MoE & 64K & vLLM & A \\
Qwen3.6-35B-A3B-FP8 & 35B/3B & MoE/FP8 & 64K & vLLM & A \\
\bottomrule
\end{tabular}
\end{table}

Models are served on two GPU systems: AMD Instinct MI300A APUs (128GB HBM3 per APU) at a shared HPC facility~\cite{mpcdf_viper} via vLLM~\cite{kwon2023vllm}, and an institutional 4$\times$ Quadro RTX 5000 system (64GB total VRAM) via llama.cpp for the Q8 variant.
The Quadro RTX 5000 based system (AMD CPU, 256GB main memory, 4$\times$16GB GPU VRAM) is vintage 2020 hardware and serves as our production deployment target.
Context windows range from 32K to 64K tokens (Table~\ref{tab:models}).
The full OWL Turtle ontology (80KB, approximately 18K tokens) and the SQL schema with comments are each provided in the system prompt, requiring models that can handle substantial context alongside the generation task.
Even the smallest context window (32K for the Q8 model) provides ample headroom, and no context overflow errors were observed in any run.
The same models are used for both SPARQL and SQL evaluation.

\subsection{Evaluation Protocol}
\label{sec:protocol}

Twenty-one competency and regression questions were developed with domain experts, covering filtered queries (``Subjects over age 42 in study 42''), multi-entity joins (``Experiments with subject from study EPEQ-7ITX''), and aggregations with sorting (``All datasets for a study, sorted by acquisition time'').
Each test case comprises an NL question, a reference SPARQL query, and expected results in SPARQL Results JSON format.
The web application's ``Save Test Case'' function generates these artifacts directly from a successful query, enabling domain experts to create new test cases without manual JSON editing.
The primary metric is \textit{result match}: a generated query is counted as correct when its result set equals the reference result set after canonicalization (row order ignored, variable names normalized).
Accuracy is the fraction of the 21 questions matched; reported times are mean LLM generation time over the questions whose generated query executed.

\textbf{SPARQL evaluation.}
Each test case is evaluated across all combinations of model, ontology representation, system prompt variant, and temperature setting: 8 models $\times$ 8 ontology representations $\times$ 3 prompts $\times$ 4 temperatures, yielding over 16,000 runs.

\textbf{SQL evaluation.}
The same test cases are evaluated across 8 models $\times$ 2 schema variants (with and without column comments) $\times$ 4 temperatures.

For both backends, when a generated query produces a syntax error, the harness sends the query and error message back to the LLM for up to two correction attempts.
As discussed in Section~\ref{sec:nlkgq}, this is technically not zero-shot, but it is rarely used with the LLMs chosen for production deployment.
LLM generation time is recorded for each run.

\subsection{Example Queries and Generated SPARQL}
\label{sec:examples}

To illustrate what the system produces, we show three competency questions of increasing complexity with the SPARQL generated by Qwen3.6-27B (default ontology, zero-shot).

\textbf{Simple lookup.}
NL: ``Show all the experiments by ID in study 101.''
\begin{lstlisting}
SELECT DISTINCT ?ExperimentId WHERE {
  ?study mro:hasStudyId "101" .
  ?experiment mro:isExperimentOfStudy ?study .
  ?experiment mro:hasExperimentId ?ExperimentId .
}
\end{lstlisting}

\textbf{Multi-entity join.}
NL: ``List all experiment IDs for study 101. Show the subject ID, handedness, age, and acquisition date for each.''
\begin{lstlisting}
SELECT DISTINCT ?ExperimentId ?SubjectId
       ?Handedness ?PatientAge ?AcquisitionDate
WHERE {
  ?exp mro:isExperimentOfStudy ?study .
  ?study mro:hasStudyId "101" .
  ?exp mro:hasExperimentId ?ExperimentId .
  ?exp mro:aExperimentHasSubject ?subj .
  ?subj mro:hasSubjectId ?SubjectId .
  ?subj mro:hasHandedness ?Handedness .
  ?exp mro:hasPatientAge ?PatientAge .
  ?exp mro:hasExperimentBeginDate ?AcquisitionDate .
}
\end{lstlisting}

\textbf{Filtered aggregation.}
NL: ``For study 42, find experiments where the subject is older than 42. List only experiments before June 2025.''
\begin{lstlisting}
SELECT DISTINCT ?ExperimentId ?PatientAge WHERE {
  ?exp mro:isExperimentOfStudy ?study .
  ?study mro:hasStudyId "42" .
  ?exp mro:hasExperimentId ?ExperimentId .
  ?exp mro:hasPatientAge ?PatientAge .
  ?exp mro:hasExperimentBeginDate ?date .
  FILTER(xsd:integer(?PatientAge) > 42)
  FILTER(?date < "2025-06-01"^^xsd:date)
}
\end{lstlisting}

All three queries are generated zero-shot from the NL question and the ontology alone.
The LLM correctly uses ontology property names, traverses relationships via inverse properties, and applies appropriate type casting in filters.
PREFIX declarations are omitted above for brevity.

Table~\ref{tab:testcases} lists all competency questions used in the evaluation.

\begin{table}[h]
\small
\caption{Competency questions by category. L=lookup, J=join, F=filter, A=aggregation, S=sort, O=ontology.}
\label{tab:testcases}
\begin{tabular}{r p{0.68\columnwidth} l}
\toprule
\# & Description & Type \\
\midrule
1 & List experiments in a study & L \\
2 & Find experiments for a subject by pseudonym & L \\
3 & Find all experiments for the subject in a given experiment & J \\
4 & Same as 3, rephrased as two-part question & J \\
5 & Experiments matching two protocol name patterns & J,F \\
6 & Experiment details with subject demographics and date & J \\
7 & Dataset names with acquisition date and time & L,S \\
8 & All properties and values on a specific dataset instance & O \\
9 & Acquisition dates for all experiments in a study, sorted & L,S \\
10 & Experiments filtered by subject age and date range & F \\
11 & Subject demographics (language, degree, handedness, etc.) & J \\
12 & Sequence datasets matching coil name substring & F \\
13 & Same as 12, exact coil name match & F \\
14 & Experiments after a time of day, sorted by duration & F,S \\
15 & Experiments in a date and duration range with study info & F,J,S \\
16 & Experiments using a specific coil with study and date & J,F \\
17 & Studies with scanner types and total experiment hours & A,J,S \\
18 & Ontology introspection: classes, properties, and types & O \\
19 & Coils used in a date range with experiment counts & A,F \\
20 & Top N coils by usage count in a year & A,F,S \\
21 & Datasets on a specific experiment, sorted by time & L,S \\
\bottomrule
\end{tabular}
\end{table}

\section{Results}
\label{sec:results}

\subsection{SPARQL: Model Comparison}
\label{sec:results_model}

Table~\ref{tab:model_results} shows the best SPARQL configuration for each model, reporting the full configuration (ontology representation, temperature, system prompt) that achieved the highest accuracy.
Short model identifiers (ID column) are used in subsequent tables.

\begin{table*}[t]
\small
\caption{Best SPARQL configuration per model (21 competency questions). The ID column provides short identifiers used in subsequent tables: Q30/Q36=Qwen generation, xxB=parameters, D=dense, M=MoE, F=FP8, Q=Q8, C=Coder. HW from Table~\ref{tab:models}.}
\label{tab:model_results}
\begin{tabular}{ll r lrl rr l}
\toprule
Model ID & Model Name & Acc\% & Ontology & Temp & Prompt & Avg Time & Tokens & HW \\
\midrule
Q36.27B.D & Qwen3.6-27B          & 100 & default & 0.0 & baseline & 8.1s & 17.6K & A \\
Q36.27B.Q & Qwen3.6-27B-Q8\_0    & 100 & default & 0.0 & procedural & 14.6s & 17.7K & N \\
Q36.27B.F & Qwen3.6-27B-FP8      & 95 & default & 0.0 & guardrails & 7.9s & 17.8K & A \\
Q30.30B.C & Qwen3-Coder-30B-A3B  & 57 & compact-grouped & 0.2 & procedural & 1.0s & 1.9K & A \\
Q36.35B.F & Qwen3.6-35B-A3B-FP8  & 57 & default & 0.6 & guardrails & 25.8s & 17.8K & A \\
Q36.35B.M & Qwen3.6-35B-A3B      & 57 & default & 0.2 & procedural & 24.2s & 17.7K & A \\
Q30.14B.D & Qwen3-14B            & 38 & default & 0.0 & procedural & 1.8s & 16.8K & A \\
Q30.08B.D & Qwen3-8B             & 24 & default & 0.6 & procedural & 0.9s & 16.8K & A \\
\bottomrule
\end{tabular}
\end{table*}

Both 27B dense variants (full-precision and Q8 quantized) achieve 100\% accuracy on the full Turtle ontology.
The Q8 model on older institutional hardware (N) is 1.8$\times$ slower but equally accurate as the full-precision model on HPC hardware (A).
The 35B MoE models peak at 57\%, well below the 27B dense models despite having more parameters, suggesting that MoE architectures may be less effective for structured generation tasks requiring precise schema adherence.
Model scale matters within architectures: accuracy improves from 24\% (8B) through 38\% (14B) to 100\% (27B dense).

\subsection{SPARQL: Ontology Representation Impact}
\label{sec:results_ontology}

Table~\ref{tab:ontology_results} shows the best configuration for each ontology representation, reporting the full configuration that achieved the highest accuracy.

\begin{table*}[t]
\small
\caption{Best SPARQL configuration per ontology representation (21 questions). Model IDs from Table~\ref{tab:model_results}.}
\label{tab:ontology_results}
\begin{tabular}{lr rl rr l}
\toprule
Ontology & Acc\% & Temp & Prompt & Avg Time & Tokens & Model ID \\
\midrule
default          & 100 & 0.0 & baseline & 8.1s & 17.6K & Q36.27B.D \\
no-comments      & 81 & 0.0 & procedural & 14.7s & 7.6K & Q36.27B.Q \\
compact-typed    & 71 & 0.0 & procedural & 3.0s & 1.9K & Q36.27B.D \\
compact-grouped  & 67 & 0.2 & procedural & 2.9s & 1.9K & Q36.27B.D \\
compact          & 62 & 0.0 & procedural & 3.1s & 1.7K & Q36.27B.D \\
raw-graph        & 48 & 0.4 & guardrails & 3.5s & 4.1K & Q36.27B.D \\
abstract-dict    & 19 & 0.6 & baseline & 20.8s & 5.7K & Q36.35B.M \\
abstract-graph   & 5 & 0.0 & guardrails & 31.8s & 3.8K & Q36.35B.F \\
\bottomrule
\end{tabular}
\end{table*}

Only \textit{default} (the complete OWL Turtle ontology with all annotations, no ablation) reaches 100\%.
Stripping \texttt{rdfs:comment} and \texttt{rdfs:label} annotations drops accuracy to 81\%.
Compact representations that retain property names achieve 62--71\%.
Abstract representations that replace readable entity names with generic identifiers collapse to 5--19\%, confirming naming as a dominant factor in accuracy.
The Tokens column shows the prompt size for each representation, ranging from 1.7K (compact) to 17.6K (default).

\textbf{Property names} are essential: the jump from \textit{compact} (62\%) to \textit{abstract-dict} (19\%) shows the LLM uses name semantics, not just graph structure.
\textbf{Annotations} matter: stripping comments costs 19 percentage points for SPARQL, and as we show below, comments are even more critical for SQL.
\textbf{Domain and range} declarations help: \textit{compact-typed} (71\%) outperforms \textit{compact} (62\%), suggesting type information reduces cross-class property confusion.

At our ontology's scale (80KB), the full Turtle fits comfortably within 32K-token context windows, so there is no practical reason to sacrifice accuracy for compactness.
Looking forward, many smaller models now support substantially larger context windows (64K to 256K tokens), suggesting these methods will scale to domains requiring more complex ontologies.

\subsection{SPARQL: System Prompt and Temperature}
\label{sec:results_prompt_temp}

Tables~\ref{tab:prompt_results} and~\ref{tab:temp_results} show the best configurations for each system prompt variant and temperature setting.

\begin{table}[b]
\small
\caption{Best SPARQL configurations per system prompt (21 questions). All use \textit{default} ontology.}
\label{tab:prompt_results}
\begin{tabular}{lr l r l}
\toprule
Prompt & Acc\% & Model ID & Temp & Avg Time \\
\midrule
baseline    & 100 & Q36.27B.D & 0.0 & 8.1s \\
\midrule
guardrails  & 100 & Q36.27B.D & 0.0 & 8.3s \\
\midrule
procedural  & 100 & Q36.27B.D & 0.0 & 8.5s \\
procedural  & 100 & Q36.27B.Q & 0.0 & 14.6s \\
\bottomrule
\end{tabular}
\end{table}

\begin{table}[b]
\small
\caption{Best SPARQL configurations per temperature (21 questions). All use \textit{default} ontology.}
\label{tab:temp_results}
\begin{tabular}{lr l l r}
\toprule
Temp & Acc\% & Model ID & Prompt & Avg Time \\
\midrule
0.0 & 100 & Q36.27B.D & baseline & 8.1s \\
0.0 & 100 & Q36.27B.D & guardrails & 8.3s \\
0.0 & 100 & Q36.27B.D & procedural & 8.5s \\
0.0 & 100 & Q36.27B.Q & procedural & 14.6s \\
\midrule
0.2 & 95 & Q36.27B.Q & baseline & 14.2s \\
\midrule
0.4 & 90 & Q36.27B.F & guardrails & 7.8s \\
\midrule
0.6 & 90 & Q36.27B.F & procedural & 8.0s \\
\bottomrule
\end{tabular}
\end{table}

Temperature and system prompt have modest effects compared to model choice and ontology representation.
All three prompts reach 100\% with the 27B dense model at $t$=0.0, and all four temperatures reach at least 90\% when paired with the right model and the \textit{default} ontology.
The \texttt{baseline} prompt (the simplest of the three) achieves 100\%, suggesting that with a well-designed ontology, elaborate prompting is unnecessary.

\subsection{Auto-SQL Results}
\label{sec:results_sql}

Table~\ref{tab:sql_results} shows auto-SQL accuracy with and without ontology-derived column comments.

\begin{table}[h]
\small
\caption{Best auto-SQL accuracy per model (21 questions, wide-table schema). Model IDs from Table~\ref{tab:model_results}.}
\label{tab:sql_results}
\begin{tabular}{l rr}
\toprule
Model ID & With comments & No comments \\
\midrule
Q36.27B.Q & 12/21 (57\%) & 9/21 (43\%) \\
Q36.27B.D & 11/21 (52\%) & 7/21 (33\%) \\
Q36.27B.F & 11/21 (52\%) & 8/21 (38\%) \\
Q36.35B.M & 11/21 (52\%) & 6/21 (29\%) \\
Q36.35B.F & 11/21 (52\%) & 7/21 (33\%) \\
Q30.14B.D &  9/21 (43\%) & 7/21 (33\%) \\
Q30.08B.D &  8/21 (38\%) & 6/21 (29\%) \\
Q30.30B.C &  8/21 (38\%) & 7/21 (33\%) \\
\bottomrule
\end{tabular}
\end{table}

The best model achieves 57\% on auto-SQL compared to 100\% on SPARQL.
Ontology-derived column comments add 3 cases for the best model (43\% without, 57\% with), a considerably larger effect than stripping annotations from the SPARQL ontology (100\% to 81\%).
The Q8 model (Q36.27B.Q) on institutional hardware achieves the highest SQL accuracy, suggesting that the chain-of-thought reasoning enabled in llama.cpp benefits the more complex SQL generation task.

\begin{table}[h]
\scriptsize
\caption{Per-query comparison using each backend's best configuration. SPARQL: Q36.27B.D, default ontology, $t$=0.0, baseline. Auto-SQL: Q36.27B.Q, with comments, $t$=0.0. Q\# from Table~\ref{tab:testcases}. Time is LLM generation in seconds.}
\label{tab:perquery}
\begin{tabular*}{\columnwidth}{@{\extracolsep{\fill}}rl cc cc}
\toprule
 & & \multicolumn{2}{c}{SPARQL} & \multicolumn{2}{c}{SQL} \\
Q\# & Query & OK & s & OK & s \\
\midrule
1  & Experiments in a study                 & \checkmark &  4 & \checkmark &  4 \\
2  & Experiments for a subject              & \checkmark &  5 & \checkmark &  4 \\
3  & All experiments for a subject          & \checkmark &  5 & \checkmark &  5 \\
4  & Same as Q3, rephrased                  & \checkmark &  6 & \checkmark &  8 \\
5  & Two protocol name patterns             & \checkmark &  9 & \checkmark &  8 \\
6  & Experiment details with demographics   & \checkmark & 10 & \checkmark &  9 \\
7  & Dataset name, acq date and time        & \checkmark &  7 & $\times$   &  6 \\
8  & All properties on a dataset            & \checkmark &  5 & $\times$   &113 \\
9  & Acquisition dates sorted               & \checkmark &  7 & $\times$   &  7 \\
10 & Age filter with date range             & \checkmark &  9 & \checkmark &  8 \\
11 & Subject demographics for a study       & \checkmark & 10 & $\times$   &  8 \\
12 & Datasets matching coil substring       & \checkmark &  9 & \checkmark & 12 \\
13 & Datasets matching exact coil           & \checkmark &  8 & \checkmark &  9 \\
14 & Experiments after 4pm, sorted          & \checkmark &  9 & $\times$   & 10 \\
15 & Experiments in duration range          & \checkmark & 12 & \checkmark & 13 \\
16 & Experiments for a coil with date       & \checkmark &  9 & $\times$   &  8 \\
17 & Studies with scanner hours             & \checkmark & 11 & $\times$   & 12 \\
18 & Ontology classes and properties        & \checkmark & 15 & $\times$   &  8 \\
19 & Coils used in a date range             & \checkmark & 10 & \checkmark & 12 \\
20 & Top N coils by experiment count        & \checkmark &  9 & $\times$   & 10 \\
21 & Datasets on experiment, sorted         & \checkmark &  6 & \checkmark &  7 \\
\midrule
   & Total                      & 21/21 &  & 12/21 & \\
\bottomrule
\end{tabular*}
\end{table}

\subsection{Practical Deployment on Modest Hardware}
\label{sec:results_practical}

For SPARQL with the \textit{default} ontology, both 27B dense variants (Q36.27B.D and Q36.27B.Q) achieve 100\% accuracy at $t$=0.0, Q36.27B.D with all three prompts, while the FP8 variant (Q36.27B.F) reaches 95\%.

The Q8-quantized model (Q36.27B.Q) running on 4$\times$ Quadro RTX 5000 GPUs (a machine several generations old) achieves 100\% SPARQL accuracy and the highest SQL accuracy (57\%), demonstrating that the approach does not require recent-generation GPUs.
This matters for institutions with privacy constraints (e.g., GDPR for human subject data) that must deploy LLMs locally.
Published benchmark estimates~\cite{llmcheck2026} indicate that Apple Silicon workstations with 128GB unified memory can run quantized 27B to 35B models at roughly 25--50 tokens per second via MLX or llama.cpp.
The low idle power consumption of these systems makes them practical for always-on deployment, and the large unified memory leaves headroom for growing ontologies requiring larger context windows.

\section{Discussion}
\label{sec:discussion}

\subsection{The Ontology as Single Source of Truth}

The central finding is that co-designing an LLM-friendly domain ontology along with the ETL logic building the KG and the domain-specific system prompt rider yields far better results than designing any one of these components independently, or worse, accepting existing components as a given.

A related finding is that ontology-driven SPARQL queries are easier for an LLM to generate accurately than an equivalent SQL query on a relational model.

Experience with a neuroimaging demonstration domain shows that end users can create well-formed natural language queries where they would not take the time to learn and write SPARQL or SQL queries.

Finally, we have found that we can run this service on local LLMs using older or less expensive hardware. It is likely that for domains without privacy constraints, cloud-based frontier LLMs with larger context windows would be faster and more capable.

The complex pipelines in prior work~\cite{zhang2024gail,zhao2025agentigraph,zhao2025cyberbot,walter2025grasp} exist for a reason: they cope with ontologies not designed for LLM consumption.
When the ontology \textit{can} be designed, which is the case for every new domain metadata project, the simpler path is available.

\subsection{SPARQL vs SQL: What the Ontology Provides}

The gap between SPARQL and SQL accuracy (100\% vs 57\%) shows that while readable naming helps both, OWL provides structural advantages that SQL DDL lacks:

\textbf{Annotations.}
Stripping \texttt{rdfs:comment} from the SPARQL ontology drops accuracy from 100\% to 81\%.
Stripping the equivalent SQL column comments drops accuracy from 57\% to 43\%.
Annotations are critical for both backends, consistent with Wretblad et al.~\cite{wretblad2024column_descriptions}.

\textbf{Domain and range declarations.}
In our conversion, \texttt{rdfs:range} determines SQL column types and \texttt{rdfs:domain} determines table placement.
These are preserved in the DDL, but the explicit class-level grouping visible in the Turtle ontology is flattened into a column list that the LLM must parse without the same semantic scaffolding.

\textbf{Inverse properties.}
In SPARQL, \texttt{owl:inverseOf} allows the LLM to traverse any relationship in either direction.
A query can use \texttt{mro:aStudyHasExperiment} or \texttt{mro:isExperimentOfStudy} interchangeably.
In SQL, there is one foreign key, and the LLM must know which table owns it and JOIN from the correct side.
The SQL system prompt must explicitly encode join directions, while the SPARQL ontology encodes them structurally.

\textbf{Entity-Attribute-Value schema.}
As an alternative to the wide-table SQL schema (one column per property, NULLs where properties are absent), we also evaluated an Entity-Attribute-Value (EAV) representation: a narrow three-column table (entity\_id, attribute, value) per class, where each row stores a single fact.
This schema was automatically derived from the same ontology.
EAV accuracy reached only 11/21 (52\%) with the best model and comments, dropping to 2/21 (10\%) without comments.
A key difficulty is that relationship attributes in EAV store prefixed entity identifiers (e.g., \texttt{study\_101}) rather than clean values (\texttt{101}) that the LLM can reason about directly.
In SPARQL, the LLM never sees internal identifiers because it operates on variables and property patterns.
In wide-table SQL, foreign keys reference clean primary keys.
EAV exposes internal naming conventions to the query generator, requiring a domain-specific prompt rider to teach the convention.
This represents a structural disadvantage that readable naming alone cannot overcome.

\textbf{Scope of the comparison.}
This result does not show that SPARQL is generally easier for LLMs to generate than SQL.
Vejvar and Fujimoto~\cite{vejvar2026spider4ssc} report the opposite ordering under terse schema linearizations, and Sequeda et al.~\cite{sequeda2023benchmark} report a KG advantage on an enterprise schema; the ordering depends on how much semantic structure each representation carries.
Both relational schemas here were derived mechanically from the ontology.
A relational schema designed by hand for these questions, with denormalized views, natural keys, and documented join paths, would likely narrow the gap, and we have not measured by how much.
What the comparison isolates is the effect of representation with data, naming, annotations, questions, and models held constant: the OWL ontology carries domain and range declarations, inverse properties, and class-level grouping as first-class structure, and a mechanical translation to DDL loses part of that scaffolding.

\subsection{Lessons Learned}

Two findings were counterintuitive:

\textbf{Qwen3 MoE variants underperformed Qwen3 dense variants.}
The 35B MoE variants peaked at 57\%, well below the 27B dense models at 100\%, despite having more total parameters.
For structured generation tasks requiring precise schema adherence, dense models appear to be the better choice at a given compute budget.

\textbf{The simplest prompt won.}
The \texttt{baseline} prompt, with the fewest instructions, achieved the highest peak accuracy.
More elaborate prompts (guardrails, procedural) performed comparably but not better, suggesting that with a well-designed ontology, the LLM needs minimal additional guidance.

Additional findings:
\textbf{Full ontology in context} consistently produced the best results across all models and configurations.
\textbf{Readable naming} had the largest measurable effect on accuracy.
\textbf{Q8 GGUF quantization} preserved accuracy completely (100\%, matching full precision); FP8 quantization showed a small drop (95\%).
\textbf{Automatic OWL-to-SQL conversion} derived a working relational schema directly from the ontology with no manual tuning.

\textbf{The test driver is essential infrastructure.}
The combinatorial test driver (Section~\ref{sec:testdriver}) is not only research tooling; it is the engine that drives ontology evolution.
The ontology, prompt rider, and competency questions co-evolve: each iteration of any of these three artifacts is validated by re-running the full suite, making regressions immediately visible.
Without this tight feedback loop, the co-design process described in Section~\ref{sec:iterative} would be more difficult. In our experience, ontology evolution slows over time but does not halt as the vocabulary and semantics are better resolved.

\subsection{Limitations}

Our evaluation covers a single metadata domain.
While the ontology design principles are domain-agnostic, generalization to other domains needs validation.
The test set of 21 competency questions, while developed with domain experts, is small; expansion is ongoing.
Because the competency questions co-evolved with the ontology and prompt rider, accuracy on novel end-user queries may differ from test set performance.
The complexity of the ontology is bounded by the available LLM context window, with some promise shown for token-dense encodings.
The volume of metadata under management is bounded by the capabilities of the SPARQL server. We use Jena Fuseki in a Docker container, but for larger archives a heavy-duty commercial solution could be used without changing the architecture.
All models are from the Qwen3 family and results may differ with other model families.
We have not compared against cloud-hosted models (GPT-4, Claude) due to institutional data privacy constraints.
The SQL baselines are auto-generated from the ontology, not designed by a database expert; the SPARQL advantage we report is specific to that setting.

\subsection{Future Work}

\begin{itemize}
    \item \textbf{Cross-domain validation}: Applying the development process and framework to additional domains. Here we are open to discussing collaboration.
    \item \textbf{Expanded MRI archive use}: The MRO ontology builds on DICOM and BIDS standards shared across MRI research sites. With site-specific ETL adaptation, the same ontology and framework could serve other neuroimaging archives.
    \item \textbf{Expanded test cases}: Growing the competency question set, including more complex query patterns.
    \item \textbf{Multi-turn refinement}: Supporting follow-up questions where the LLM refines a previous query based on user feedback, making the web application more conversational.
    \item \textbf{LLM Model diversity}: Evaluating non-Qwen model families and cloud-hosted models where domain privacy constraints permit.
\end{itemize}

\section{Conclusion}
\label{sec:conclusion}

We have presented a reusable framework and development process for natural language access to domain-specific metadata.
A key insight is that capturing domain vocabulary and semantics in a well-designed OWL ontology enables LLM-driven query generation against both SPARQL and SQL backends, with no fine-tuning, no retrieval augmentation, and no multi-agent orchestration.

The ontology serves as the single source of truth: it defines the domain vocabulary, informs the ETL pipeline, and provides the LLM with schema context for SPARQL queries.
To explore SPARQL vs SQL for this work, we automatically generate a relational schema, transform and load the KG into a PostgreSQL database, and evaluate the competency questions via SQL, enabling a direct comparison.
On our MRI neuroimaging metadata benchmark, NL text-to-SPARQL achieves 100\% accuracy on our competency/regression question set while the analogous NL text-to-SQL achieves 57\%.
An ablation study across eight ontology representations confirms that ontology design, particularly readable naming and semantic annotations, is the dominant factor in accuracy.

The framework and development process are intended to be reusable.
The framework is designed to avoid domain-specific components.
The iterative development process outlined here (ontology, competency questions, and prompt rider evolving together) requires domain expertise but no machine learning, database, or programming skills.
We have demonstrated the process and deployed the framework at a major neuroscience
institute enabling end-user natural language search of MRI metadata on a growing image archive.

We intend to open source this work including the MRI metadata domain and the framework, and we welcome collaboration on application to new domains.

\section*{GenAI Usage Disclosure}

Generative AI tools were used to assist with manuscript editing and \LaTeX{} formatting.
Generative AI was also used to assist with coding, particularly the web server where we have little expertise available.
The system described in this paper uses locally deployed LLMs (Qwen3 family) for SPARQL and SQL query generation, which is the subject of this research.
All scientific content, experimental design, analysis, and conclusions are the work of the authors.

\bibliographystyle{ACM-Reference-Format}
\bibliography{references}


\begin{thebibliography}{35}


\ifx \showCODEN    \undefined \def \showCODEN     #1{\unskip}     \fi
\ifx \showISBNx    \undefined \def \showISBNx     #1{\unskip}     \fi
\ifx \showISBNxiii \undefined \def \showISBNxiii  #1{\unskip}     \fi
\ifx \showISSN     \undefined \def \showISSN      #1{\unskip}     \fi
\ifx \showLCCN     \undefined \def \showLCCN      #1{\unskip}     \fi
\ifx \shownote     \undefined \def \shownote      #1{#1}          \fi
\ifx \showarticletitle \undefined \def \showarticletitle #1{#1}   \fi
\ifx \showURL      \undefined \def \showURL       {\relax}        \fi
\providecommand\bibfield[2]{#2}
\providecommand\bibinfo[2]{#2}
\providecommand\natexlab[1]{#1}
\providecommand\showeprint[2][]{arXiv:#2}

\bibitem[Arazzi et~al\mbox{.}(2025)]%
        {arazzi2025sparqllm}
\bibfield{author}{\bibinfo{person}{Marco Arazzi}, \bibinfo{person}{Davide
  Ligari}, \bibinfo{person}{Serena Nicolazzo}, {and} \bibinfo{person}{Antonino
  Nocera}.} \bibinfo{year}{2025}\natexlab{}.
\newblock \bibinfo{title}{Augmented Knowledge Graph Querying leveraging
  {LLM}s}.
\newblock \bibinfo{howpublished}{arXiv preprint arXiv:2502.01298}.
\newblock


\bibitem[Carroll et~al\mbox{.}(2004)]%
        {carroll2004jena}
\bibfield{author}{\bibinfo{person}{Jeremy~J. Carroll}, \bibinfo{person}{Ian
  Dickinson}, \bibinfo{person}{Chris Dollin}, \bibinfo{person}{Dave Reynolds},
  \bibinfo{person}{Andy Seaborne}, {and} \bibinfo{person}{Kevin Wilkinson}.}
  \bibinfo{year}{2004}\natexlab{}.
\newblock \showarticletitle{Jena: Implementing the Semantic Web
  Recommendations}. In \bibinfo{booktitle}{\emph{Proceedings of the 13th
  International World Wide Web Conference (WWW)}}. \bibinfo{publisher}{ACM},
  \bibinfo{pages}{74--83}.
\newblock
\href{https://doi.org/10.1145/1013367.1013381}{doi:\nolinkurl{10.1145/1013367.1013381}}


\bibitem[Chen et~al\mbox{.}(2025)]%
        {chen2024beaver}
\bibfield{author}{\bibinfo{person}{Peter~Baile Chen}, \bibinfo{person}{Fabian
  Wenz}, \bibinfo{person}{Yi Zhang}, \bibinfo{person}{Devin Yang},
  \bibinfo{person}{Justin Choi}, \bibinfo{person}{Nesime Tatbul},
  \bibinfo{person}{Michael Cafarella}, \bibinfo{person}{{\c{C}}a{\u{g}}atay
  Demiralp}, {and} \bibinfo{person}{Michael Stonebraker}.}
  \bibinfo{year}{2025}\natexlab{}.
\newblock \showarticletitle{{BEAVER}: An Enterprise Benchmark for
  Text-to-{SQL}}.
\newblock \bibinfo{journal}{\emph{arXiv preprint arXiv:2409.02038v2}}
  (\bibinfo{year}{2025}).
\newblock


\bibitem[D'Abramo et~al\mbox{.}(2025)]%
        {dabramo2025icl_sparql}
\bibfield{author}{\bibinfo{person}{Jacopo D'Abramo}, \bibinfo{person}{Andrea
  Zugarini}, {and} \bibinfo{person}{Paolo Torroni}.}
  \bibinfo{year}{2025}\natexlab{}.
\newblock \showarticletitle{Investigating Large Language Models for
  Text-to-SPARQL Generation}. In \bibinfo{booktitle}{\emph{Proceedings of the
  4th International Workshop on Knowledge-Augmented Methods for NLP
  (KnowledgeNLP)}}. \bibinfo{publisher}{Association for Computational
  Linguistics}, \bibinfo{pages}{66--80}.
\newblock


\bibitem[Fitch et~al\mbox{.}(2022)]%
        {fitch2022mrdata}
\bibfield{author}{\bibinfo{person}{Blake~G. Fitch}, \bibinfo{person}{Sebastian
  M{\"u}ller}, {and} \bibinfo{person}{Dario Bosch}.}
  \bibinfo{year}{2022}\natexlab{}.
\newblock \showarticletitle{{MrData}: An {iRODS} Based Human Research Data
  Management System}. In \bibinfo{booktitle}{\emph{Proceedings of the iRODS
  User Group Meeting}}. \bibinfo{address}{Leuven, Belgium}.
\newblock


\bibitem[Giuliani et~al\mbox{.}(2026)]%
        {giuliani2026sparql}
\bibfield{author}{\bibinfo{person}{Alessandro Giuliani},
  \bibinfo{person}{Marco~Manolo Manca}, \bibinfo{person}{Leonardo Piano},
  \bibinfo{person}{Alessandro~Sebastian Podda}, \bibinfo{person}{Livio
  Pompianu}, {and} \bibinfo{person}{Sandro~Gabriele Tiddia}.}
  \bibinfo{year}{2026}\natexlab{}.
\newblock \showarticletitle{Are {LLM}s adequate {SPARQL} query generators?
  {I}nvestigating zero-shot {NL}-to-{SPARQL} translation}.
\newblock \bibinfo{journal}{\emph{Neural Computing and Applications}}
  \bibinfo{volume}{38}, Article \bibinfo{articleno}{33} (\bibinfo{date}{feb}
  \bibinfo{year}{2026}), \bibinfo{numpages}{38}~pages.
\newblock
\href{https://doi.org/10.1007/s00521-025-11799-x}{doi:\nolinkurl{10.1007/s00521-025-11799-x}}


\bibitem[Gorgolewski et~al\mbox{.}(2016)]%
        {gorgolewski2016bids}
\bibfield{author}{\bibinfo{person}{Krzysztof~J Gorgolewski} {et~al\mbox{.}}}
  \bibinfo{year}{2016}\natexlab{}.
\newblock \showarticletitle{The brain imaging data structure, a format for
  organizing and describing outputs of neuroimaging experiments}.
\newblock \bibinfo{journal}{\emph{Scientific Data}}  \bibinfo{volume}{3}
  (\bibinfo{year}{2016}), \bibinfo{pages}{160044}.
\newblock
\href{https://doi.org/10.1038/sdata.2016.44}{doi:\nolinkurl{10.1038/sdata.2016.44}}


\bibitem[Hogan et~al\mbox{.}(2021)]%
        {hogan2021kg}
\bibfield{author}{\bibinfo{person}{Aidan Hogan}, \bibinfo{person}{Eva
  Blomqvist}, \bibinfo{person}{Michael Cochez}, \bibinfo{person}{Claudia
  d'Amato}, \bibinfo{person}{Gerard de Melo}, \bibinfo{person}{Claudio
  Gutierrez}, \bibinfo{person}{Sabrina Kirrane}, \bibinfo{person}{Jose
  Emilio~Labra Gayo}, \bibinfo{person}{Roberto Navigli},
  \bibinfo{person}{Sebastian Neumaier}, {et~al\mbox{.}}}
  \bibinfo{year}{2021}\natexlab{}.
\newblock \showarticletitle{Knowledge Graphs}.
\newblock \bibinfo{journal}{\emph{Comput. Surveys}} \bibinfo{volume}{54},
  \bibinfo{number}{4} (\bibinfo{year}{2021}), \bibinfo{pages}{1--37}.
\newblock
\href{https://doi.org/10.1145/3447772}{doi:\nolinkurl{10.1145/3447772}}


\bibitem[Kosten et~al\mbox{.}(2023)]%
        {kosten2023spider4sparql}
\bibfield{author}{\bibinfo{person}{Catherine Kosten}, \bibinfo{person}{Philippe
  Cudr{\'e}-Mauroux}, {and} \bibinfo{person}{Kurt Stockinger}.}
  \bibinfo{year}{2023}\natexlab{}.
\newblock \showarticletitle{{Spider4SPARQL}: A Complex Benchmark for Evaluating
  Knowledge Graph Question Answering Systems}. In
  \bibinfo{booktitle}{\emph{2023 IEEE International Conference on Big Data
  (BigData)}}. \bibinfo{publisher}{IEEE}, \bibinfo{pages}{5272--5281}.
\newblock


\bibitem[Kwon et~al\mbox{.}(2023)]%
        {kwon2023vllm}
\bibfield{author}{\bibinfo{person}{Woosuk Kwon}, \bibinfo{person}{Zhuohan Li},
  \bibinfo{person}{Siyuan Zhuang}, \bibinfo{person}{Ying Sheng},
  \bibinfo{person}{Lianmin Zheng}, \bibinfo{person}{Cody~Hao Yu},
  \bibinfo{person}{Joseph~E Gonzalez}, \bibinfo{person}{Hao Zhang}, {and}
  \bibinfo{person}{Ion Stoica}.} \bibinfo{year}{2023}\natexlab{}.
\newblock \showarticletitle{Efficient Memory Management for Large Language
  Model Serving with {PagedAttention}}. In
  \bibinfo{booktitle}{\emph{Proceedings of SOSP}}.
\newblock
\href{https://doi.org/10.1145/3600006.3613165}{doi:\nolinkurl{10.1145/3600006.3613165}}


\bibitem[Li et~al\mbox{.}(2016)]%
        {li2016dcm2niix}
\bibfield{author}{\bibinfo{person}{Xiangrui Li}, \bibinfo{person}{Paul~S
  Morgan}, \bibinfo{person}{John Ashburner}, \bibinfo{person}{Jolinda Smith},
  {and} \bibinfo{person}{Christopher Rorden}.} \bibinfo{year}{2016}\natexlab{}.
\newblock \showarticletitle{The first step for neuroimaging data analysis:
  {DICOM} to {NIfTI} conversion}.
\newblock \bibinfo{journal}{\emph{Journal of Neuroscience Methods}}
  \bibinfo{volume}{264} (\bibinfo{year}{2016}), \bibinfo{pages}{47--56}.
\newblock
\href{https://doi.org/10.1016/j.jneumeth.2016.03.001}{doi:\nolinkurl{10.1016/j.jneumeth.2016.03.001}}


\bibitem[{LLM Check}(2026)]%
        {llmcheck2026}
\bibfield{author}{\bibinfo{person}{{LLM Check}}.}
  \bibinfo{year}{2026}\natexlab{}.
\newblock \bibinfo{title}{Apple Silicon {LLM} Benchmarks: Real tok/s by Model,
  Chip and Quantization}.
\newblock \bibinfo{howpublished}{\url{https://llmcheck.net/benchmarks}}.
\newblock


\bibitem[Mader and Kleemeyer(2023)]%
        {mader2023castellum}
\bibfield{author}{\bibinfo{person}{Karolina Mader} {and} \bibinfo{person}{Maike
  Kleemeyer}.} \bibinfo{year}{2023}\natexlab{}.
\newblock \showarticletitle{Castellum: A Data Protection-Compliant Web
  Application for the Subject Management of Human Science Studies}. In
  \bibinfo{booktitle}{\emph{Proceedings of the Conference on Research Data
  Infrastructure (CoRDI)}}, Vol.~\bibinfo{volume}{1}.
\newblock
\href{https://doi.org/10.52825/cordi.v1i.325}{doi:\nolinkurl{10.52825/cordi.v1i.325}}


\bibitem[{Max Planck Computing and Data Facility}(2025)]%
        {mpcdf_viper}
\bibfield{author}{\bibinfo{person}{{Max Planck Computing and Data Facility}}.}
  \bibinfo{year}{2025}\natexlab{}.
\newblock \bibinfo{title}{Viper-{GPU} User Guide}.
\newblock
  \bibinfo{howpublished}{\url{https://docs.mpcdf.mpg.de/doc/computing/viper-gpu-user-guide.html}}.
\newblock
\newblock
\shownote{228 nodes, 2$\times$ AMD Instinct MI300A APUs per node, 128\,GB HBM3
  per APU}.


\bibitem[Noy and McGuinness(2001)]%
        {noy2001ontology101}
\bibfield{author}{\bibinfo{person}{Natalya~F. Noy} {and}
  \bibinfo{person}{Deborah~L. McGuinness}.} \bibinfo{year}{2001}\natexlab{}.
\newblock \bibinfo{booktitle}{\emph{Ontology Development 101: A Guide to
  Creating Your First Ontology}}.
\newblock \bibinfo{type}{{T}echnical {R}eport}. \bibinfo{institution}{Stanford
  University}.
\newblock
\newblock
\shownote{Stanford Knowledge Systems Laboratory Technical Report KSL-01-05}.


\bibitem[Paslavska et~al\mbox{.}(2026)]%
        {paslavska2026rhizomerllm}
\bibfield{author}{\bibinfo{person}{D. Paslavska}, \bibinfo{person}{J.M.
  L{\'o}pez-Gil}, \bibinfo{person}{J. Pereira}, \bibinfo{person}{R. Gil}, {and}
  \bibinfo{person}{R. Garc{\'i}a}.} \bibinfo{year}{2026}\natexlab{}.
\newblock \showarticletitle{{Rhizomer-LLM}: Natural language interface for
  semantic data exploration}.
\newblock \bibinfo{journal}{\emph{SoftwareX}}  \bibinfo{volume}{35}
  (\bibinfo{year}{2026}), \bibinfo{pages}{102834}.
\newblock
\href{https://doi.org/10.1016/j.softx.2026.102834}{doi:\nolinkurl{10.1016/j.softx.2026.102834}}


\bibitem[Rajkumar et~al\mbox{.}(2022)]%
        {rajkumar2022schema}
\bibfield{author}{\bibinfo{person}{Nitarshan Rajkumar},
  \bibinfo{person}{Raymond Li}, {and} \bibinfo{person}{Dzmitry Bahdanau}.}
  \bibinfo{year}{2022}\natexlab{}.
\newblock \showarticletitle{Evaluating the Text-to-{SQL} Capabilities of Large
  Language Models}.
\newblock \bibinfo{journal}{\emph{arXiv preprint arXiv:2204.00498}}
  (\bibinfo{year}{2022}).
\newblock


\bibitem[Rasheed and Aguado(2025)]%
        {rasheed2025domain_sparql}
\bibfield{author}{\bibinfo{person}{Mohammed~H. Rasheed} {and}
  \bibinfo{person}{Marina Aguado}.} \bibinfo{year}{2025}\natexlab{}.
\newblock \showarticletitle{{LLM}-Based Natural Language to {SPARQL}
  Translation over Domain-Specific Knowledge Graph}.
\newblock \bibinfo{journal}{\emph{Knowledge Organization}}
  \bibinfo{volume}{52}, \bibinfo{number}{8} (\bibinfo{year}{2025}),
  \bibinfo{pages}{42705}.
\newblock
\href{https://doi.org/10.31083/KO42705}{doi:\nolinkurl{10.31083/KO42705}}


\bibitem[{RDFLib Team}(2024)]%
        {rdflib}
\bibfield{author}{\bibinfo{person}{{RDFLib Team}}.}
  \bibinfo{year}{2024}\natexlab{}.
\newblock \bibinfo{title}{rdflib: A Python library for working with {RDF}}.
\newblock \bibinfo{howpublished}{\url{https://github.com/RDFLib/rdflib}}.
\newblock


\bibitem[Rony et~al\mbox{.}(2022)]%
        {rony2022sgpt}
\bibfield{author}{\bibinfo{person}{Md~Rashad Al~Hasan Rony},
  \bibinfo{person}{Uttam Kumar}, \bibinfo{person}{Roman Teucher},
  \bibinfo{person}{Liubov Kovriguina}, {and} \bibinfo{person}{Jens Lehmann}.}
  \bibinfo{year}{2022}\natexlab{}.
\newblock \showarticletitle{{SGPT}: A Generative Approach for {SPARQL} Query
  Generation From Natural Language Questions}.
\newblock \bibinfo{journal}{\emph{IEEE Access}}  \bibinfo{volume}{10}
  (\bibinfo{year}{2022}), \bibinfo{pages}{70712--70723}.
\newblock
\href{https://doi.org/10.1109/ACCESS.2022.3188714}{doi:\nolinkurl{10.1109/ACCESS.2022.3188714}}


\bibitem[Sequeda et~al\mbox{.}(2023)]%
        {sequeda2023benchmark}
\bibfield{author}{\bibinfo{person}{Juan~F. Sequeda}, \bibinfo{person}{Dean
  Allemang}, {and} \bibinfo{person}{Bryon Jacob}.}
  \bibinfo{year}{2023}\natexlab{}.
\newblock \showarticletitle{A Benchmark to Understand the Role of Knowledge
  Graphs on Large Language Model's Accuracy for Question Answering on
  Enterprise {SQL} Databases}.
\newblock \bibinfo{journal}{\emph{arXiv preprint arXiv:2311.07509}}
  (\bibinfo{year}{2023}).
\newblock


\bibitem[Soru et~al\mbox{.}(2017)]%
        {soru2017sparql}
\bibfield{author}{\bibinfo{person}{Tommaso Soru}, \bibinfo{person}{Edgard
  Marx}, \bibinfo{person}{Diego Moussallem}, \bibinfo{person}{Gustavo Publio},
  \bibinfo{person}{Andr{\'e} Valdestilhas}, \bibinfo{person}{Diego Esteves},
  {and} \bibinfo{person}{Ciro~Baron Neto}.} \bibinfo{year}{2017}\natexlab{}.
\newblock \showarticletitle{{SPARQL} as a Foreign Language}. In
  \bibinfo{booktitle}{\emph{SEMANTiCS 2017 Posters \& Demos}}.
\newblock


\bibitem[Usbeck et~al\mbox{.}(2018)]%
        {usbeck2018qald9}
\bibfield{author}{\bibinfo{person}{Ricardo Usbeck}, \bibinfo{person}{Ria~Hari
  Gusmita}, \bibinfo{person}{Axel-Cyrille~Ngonga Ngomo}, {and}
  \bibinfo{person}{Muhammad Saleem}.} \bibinfo{year}{2018}\natexlab{}.
\newblock \showarticletitle{9th Challenge on Question Answering over Linked
  Data ({QALD-9})}. In \bibinfo{booktitle}{\emph{Joint Proceedings of SemDeep-4
  and NLIWoD-4 and QALD-9, co-located with ISWC 2018}}
  \emph{(\bibinfo{series}{CEUR Workshop Proceedings},
  Vol.~\bibinfo{volume}{2241})}. \bibinfo{pages}{58--64}.
\newblock


\bibitem[Vejvar and Fujimoto(2026)]%
        {vejvar2026spider4ssc}
\bibfield{author}{\bibinfo{person}{Martin Vejvar} {and}
  \bibinfo{person}{Yasutaka Fujimoto}.} \bibinfo{year}{2026}\natexlab{}.
\newblock \showarticletitle{Are we too focused on single query language?
  {I}nvestigating text-to-{SQL}/{SPARQL}/{C}ypher task complexity via
  fine-tuning unbiased {T5} models}.
\newblock \bibinfo{journal}{\emph{Neurocomputing}}  \bibinfo{volume}{697}
  (\bibinfo{year}{2026}), \bibinfo{pages}{134202}.
\newblock
\href{https://doi.org/10.1016/j.neucom.2026.134202}{doi:\nolinkurl{10.1016/j.neucom.2026.134202}}


\bibitem[Vrande{\v{c}}i{\'c} and Kr{\"o}tzsch(2014)]%
        {vrandecic2014wikidata}
\bibfield{author}{\bibinfo{person}{Denny Vrande{\v{c}}i{\'c}} {and}
  \bibinfo{person}{Markus Kr{\"o}tzsch}.} \bibinfo{year}{2014}\natexlab{}.
\newblock \showarticletitle{Wikidata: A Free Collaborative Knowledgebase}.
\newblock \bibinfo{journal}{\emph{Commun. ACM}} \bibinfo{volume}{57},
  \bibinfo{number}{10} (\bibinfo{year}{2014}), \bibinfo{pages}{78--85}.
\newblock
\href{https://doi.org/10.1145/2629489}{doi:\nolinkurl{10.1145/2629489}}


\bibitem[{W3C}(2012)]%
        {owl2}
\bibfield{author}{\bibinfo{person}{{W3C}}.} \bibinfo{year}{2012}\natexlab{}.
\newblock \bibinfo{title}{{OWL} 2 Web Ontology Language Primer (Second
  Edition)}.
\newblock \bibinfo{howpublished}{\url{https://www.w3.org/TR/owl2-primer/}}.
\newblock


\bibitem[{W3C}(2013)]%
        {sparql11}
\bibfield{author}{\bibinfo{person}{{W3C}}.} \bibinfo{year}{2013}\natexlab{}.
\newblock \bibinfo{title}{{SPARQL} 1.1 Query Language}.
\newblock \bibinfo{howpublished}{\url{https://www.w3.org/TR/sparql11-query/}}.
\newblock


\bibitem[Walter and Bast(2025)]%
        {walter2025grasp}
\bibfield{author}{\bibinfo{person}{Sebastian Walter} {and}
  \bibinfo{person}{Hannah Bast}.} \bibinfo{year}{2025}\natexlab{}.
\newblock \showarticletitle{{GRASP}: Generic Reasoning And {SPARQL} Generation
  across Knowledge Graphs}. In \bibinfo{booktitle}{\emph{The Semantic Web --
  ISWC 2025}} \emph{(\bibinfo{series}{Lecture Notes in Computer Science},
  Vol.~\bibinfo{volume}{16140})}. \bibinfo{publisher}{Springer}.
\newblock
\href{https://doi.org/10.1007/978-3-032-09527-5_15}{doi:\nolinkurl{10.1007/978-3-032-09527-5_15}}


\bibitem[Wilkinson et~al\mbox{.}(2016)]%
        {wilkinson2016fair}
\bibfield{author}{\bibinfo{person}{Mark~D Wilkinson} {et~al\mbox{.}}}
  \bibinfo{year}{2016}\natexlab{}.
\newblock \showarticletitle{The {FAIR} Guiding Principles for scientific data
  management and stewardship}.
\newblock \bibinfo{journal}{\emph{Scientific Data}}  \bibinfo{volume}{3}
  (\bibinfo{year}{2016}), \bibinfo{pages}{160018}.
\newblock


\bibitem[Wretblad et~al\mbox{.}(2024)]%
        {wretblad2024column_descriptions}
\bibfield{author}{\bibinfo{person}{Niklas Wretblad}, \bibinfo{person}{Oskar
  Holmstr{\"o}m}, \bibinfo{person}{Erik Larsson}, \bibinfo{person}{Axel
  Wiks{\"a}ter}, \bibinfo{person}{Oscar S{\"o}derlund},
  \bibinfo{person}{Hjalmar {\"O}hman}, \bibinfo{person}{Ture Pont{\'e}n},
  \bibinfo{person}{Martin Forsberg}, \bibinfo{person}{Martin S{\"o}rme}, {and}
  \bibinfo{person}{Fredrik Heintz}.} \bibinfo{year}{2024}\natexlab{}.
\newblock \showarticletitle{Synthetic {SQL} Column Descriptions and Their
  Impact on Text-to-{SQL} Performance}.
\newblock \bibinfo{journal}{\emph{arXiv preprint arXiv:2408.04691}}
  (\bibinfo{year}{2024}).
\newblock


\bibitem[Yu et~al\mbox{.}(2018)]%
        {yu2018spider}
\bibfield{author}{\bibinfo{person}{Tao Yu}, \bibinfo{person}{Rui Zhang},
  \bibinfo{person}{Kai Yang}, \bibinfo{person}{Michihiro Yasunaga},
  \bibinfo{person}{Dongxu Wang}, \bibinfo{person}{Zifan Li},
  \bibinfo{person}{James Ma}, \bibinfo{person}{Irene Li},
  \bibinfo{person}{Qingning Yao}, \bibinfo{person}{Shanelle Roman},
  {et~al\mbox{.}}} \bibinfo{year}{2018}\natexlab{}.
\newblock \showarticletitle{Spider: A Large-Scale Human-Labeled Dataset for
  Complex and Cross-Domain Semantic Parsing and Text-to-{SQL} Task}. In
  \bibinfo{booktitle}{\emph{Proceedings of EMNLP}}.
\newblock


\bibitem[Zahera et~al\mbox{.}(2024)]%
        {zahera2024cotsparql}
\bibfield{author}{\bibinfo{person}{Hamada~M. Zahera}, \bibinfo{person}{Manzoor
  Ali}, \bibinfo{person}{Mohamed~Ahmed Sherif}, \bibinfo{person}{Diego
  Moussallem}, {and} \bibinfo{person}{Axel-Cyrille Ngonga~Ngomo}.}
  \bibinfo{year}{2024}\natexlab{}.
\newblock \showarticletitle{Generating {SPARQL} from Natural Language Using
  Chain-of-Thoughts Prompting}.
\newblock In \bibinfo{booktitle}{\emph{Knowledge Graphs in the Age of Language
  Models and Neuro-Symbolic AI (SEMANTiCS 2024)}}. \bibinfo{series}{Studies on
  the Semantic Web}, Vol.~\bibinfo{volume}{60}. \bibinfo{publisher}{IOS Press},
  \bibinfo{pages}{353--368}.
\newblock
\href{https://doi.org/10.3233/SSW240028}{doi:\nolinkurl{10.3233/SSW240028}}


\bibitem[Zhang et~al\mbox{.}(2024)]%
        {zhang2024gail}
\bibfield{author}{\bibinfo{person}{Zhiqiang Zhang}, \bibinfo{person}{Liqiang
  Wen}, {and} \bibinfo{person}{Wen Zhao}.} \bibinfo{year}{2024}\natexlab{}.
\newblock \showarticletitle{A {GAIL} Fine-Tuned {LLM} Enhanced Framework for
  Low-Resource Knowledge Graph Question Answering}. In
  \bibinfo{booktitle}{\emph{Proceedings of the 33rd ACM International
  Conference on Information and Knowledge Management (CIKM)}}.
  \bibinfo{publisher}{ACM}, \bibinfo{pages}{3300--3309}.
\newblock
\href{https://doi.org/10.1145/3627673.3679753}{doi:\nolinkurl{10.1145/3627673.3679753}}


\bibitem[Zhao et~al\mbox{.}(2025b)]%
        {zhao2025cyberbot}
\bibfield{author}{\bibinfo{person}{Chengshuai Zhao}, \bibinfo{person}{Riccardo
  De~Maria}, \bibinfo{person}{Tharindu Kumarage}, \bibinfo{person}{Kumar~Satvik
  Chaudhary}, \bibinfo{person}{Garima Agrawal}, \bibinfo{person}{Yiwen Li},
  \bibinfo{person}{Jongchan Park}, \bibinfo{person}{Yuli Deng},
  \bibinfo{person}{Ying-Chih Chen}, {and} \bibinfo{person}{Huan Liu}.}
  \bibinfo{year}{2025}\natexlab{b}.
\newblock \showarticletitle{{CyberBOT}: Ontology-Grounded Retrieval Augmented
  Generation for Reliable Cybersecurity Education}. In
  \bibinfo{booktitle}{\emph{Proceedings of the 34th ACM International
  Conference on Information and Knowledge Management (CIKM)}}.
  \bibinfo{publisher}{ACM}.
\newblock
\href{https://doi.org/10.1145/3746252.3761478}{doi:\nolinkurl{10.1145/3746252.3761478}}


\bibitem[Zhao et~al\mbox{.}(2025a)]%
        {zhao2025agentigraph}
\bibfield{author}{\bibinfo{person}{Xinjie Zhao}, \bibinfo{person}{Moritz Blum},
  \bibinfo{person}{Fan Gao}, \bibinfo{person}{Yingjian Chen},
  \bibinfo{person}{Boming Yang}, \bibinfo{person}{Luis Marquez-Carpintero},
  \bibinfo{person}{M{\'o}nica Pina-Navarro}, \bibinfo{person}{Yanran Fu},
  \bibinfo{person}{So Morikawa}, \bibinfo{person}{Yusuke Iwasawa},
  \bibinfo{person}{Yutaka Matsuo}, \bibinfo{person}{Chanjun Park}, {and}
  \bibinfo{person}{Irene Li}.} \bibinfo{year}{2025}\natexlab{a}.
\newblock \showarticletitle{{AGENTiGraph}: A Multi-Agent Knowledge Graph
  Framework for Interactive, Domain-Specific {LLM} Chatbots}. In
  \bibinfo{booktitle}{\emph{Proceedings of the 34th ACM International
  Conference on Information and Knowledge Management (CIKM)}}.
  \bibinfo{publisher}{ACM}.
\newblock
\href{https://doi.org/10.1145/3746252.3761459}{doi:\nolinkurl{10.1145/3746252.3761459}}


\end{thebibliography}

\end{document}